\documentstyle[12pt]{article}
\def\PRL #1 #2 #3{{\sl Phys. Rev. Lett.} {\bf#1} (#2) #3}
\def\NPB #1 #2 #3{{\sl Nucl. Phys.} {\bf B #1} (#2) #3}
\def\NPBFS #1 #2 #3 #4{{\sl Nucl. Phys.} {\bf B #2} [FS#1] (#3) #4}
\def\CMP #1 #2 #3{{\sl Commun. Math. Phys.} {\bf #1} (#2) #3}
\def\PRD #1 #2 #3{{\sl Phys. Rev.} {\bf D #1} (#2) #3}
\def\PLA #1 #2 #3{{\sl Phys. Lett.} {\bf A #1} (#2) #3}
\def\PLB #1 #2 #3{{\sl Phys. Lett.} {\bf B #1} (#2) #3}
\def\JMP #1 #2 #3{{\sl J. Math. Phys.} {\bf #1} (#2) #3}
\def\PTP #1 #2 #3{{\sl Prog. Theor. Phys.} {\bf #1} (#2) #3}
\def\SPTP #1 #2 #3{{\sl Suppl. Prog. Theor. Phys.} {\bf #1} (#2) #3}
\def\AoP #1 #2 #3{{\sl Ann. of Phys.} {\bf #1} (#2) #3}
\def\PNAS #1 #2 #3{{\sl Proc. Natl. Acad. Sci. USA} {\bf #1} (#2) #3}
\def\RMP #1 #2 #3{{\sl Rev. Mod. Phys.} {\bf #1} (#2) #3}
\def\PR #1 #2 #3{{\sl Phys. Reports} {\bf #1} (#2) #3}
\def\AoM #1 #2 #3{{\sl Ann. of Math.} {\bf #1} (#2) #3}
\def\UMN #1 #2 #3{{\sl Usp. Mat. Nauk} {\bf #1} (#2) #3}
\def\FAP #1 #2 #3{{\sl Funkt. Anal. Prilozheniya} {\bf #1} (#2) #3}
\def\FAaIA #1 #2 #3{{\sl Functional Analysis and Its Application} {\bf
#1} (#2) #3}
\def\BAMS #1 #2 #3{{\sl Bull. Am. Math. Soc.} {\bf #1} (#2)
#3} \def\TAMS #1 #2 #3{{\sl Trans. Am. Math. Soc.} {\bf #1} (#2) #3}
\def\InvM #1 #2 #3{{\sl Invent. Math.} {\bf #1} (#2) #3}
\def\LMP #1 #2 #3{{\sl Letters in Math. Phys.} {\bf #1} (#2) #3}
\def\IJMPA #1 #2 #3{{\sl Int. J. Mod. Phys.} {\bf A #1} (#2) #3}
\def\AdM #1 #2 #3{{\sl Advances in Math.} {\bf #1} (#2) #3}
\def\RMaP #1 #2 #3{{\sl Reports on Math. Phys.} {\bf #1} (#2) #3}
\def\IJM #1 #2 #3{{\sl Ill. J. Math.} {\bf #1} (#2) #3}
\def\APP #1 #2 #3{{\sl Acta Phys. Polon.} {\bf #1} (#2) #3}
\def\TMP #1 #2 #3{{\sl Theor. Mat. Phys.} {\bf #1} (#2) #3}
\def\JPA #1 #2 #3{{\sl J. Physics} {\bf A#1} (#2) #3}
\def\JSM #1 #2 #3{{\sl J. Soviet Math.} {\bf #1} (#2) #3}
\def\MPLA #1 #2 #3{{\sl Mod. Phys. Lett.} {\bf A #1} (#2) #3}
\def\JETP #1 #2 #3{{\sl Sov. Phys. JETP} {\bf #1} (#2) #3}
\def\JETPL #1 #2 #3{{\sl  Sov. Phys. JETP Lett.} {\bf #1} (#2) #3}
\def\PHSA #1 #2 #3{{\sl Physica} {\bf A #1} (#2) #3}
\def\CQG #1 #2 #3{{\sl Class. Quantum Grav.} {\bf #1} (#2) #3}
\def\SJNP #1 #2 #3{{\sl Sov. J. Nucl. Phys. (Yadern.Fiz.)} 
{\bf #1} (#2) #3}
\def\a{\alpha}\def\b{\beta}\def\g{\gamma}\def\d{\delta}\def\e{\epsilon}

\def\l{\lambda}\def\L{\Lambda}
\def\Th{\Theta}\def\th{\theta}\def\Om{\Omega}

\newcommand{\nn}{\nonumber\\}\newcommand{\p}[1]{(\ref{#1})}
\begin{document}
\renewcommand{\thefootnote}{\fnsymbol{footnote}}
\thispagestyle{empty}
\begin{flushright}
Preprint DFPD 96/TH/15\\
hep-th/9603187\\
\end{flushright}

\medskip
\begin{center}
{\large\bf Hamiltonian Reduction of 
Supersymmetric WZNW}\\
{\large\bf Models on Bosonic Groups
and Superstrings
}
\vspace{0.3cm}

{\bf Dmitri Sorokin\footnote{On leave from Kharkov Institute of
Physics and Technology, Kharkov, 310108, Ukraine.}
and Francesco Toppan,}\\
\vspace{0.2cm}
{\it Dipartimento di Fisica ``Galileo Galilei'',\\
Universit\`a degli Studi di Padova\\
and $~^{(\star )}$ INFN, Sezione di Padova;\\
Via F. Marzolo 8, 35131 Padova, Italy.}\\
e--mail: sorokin@pd.infn.it\\
~~~~~~~~~~toppan@pd.infn.it\\ 

\vspace{0.5cm}
{\bf Abstract}
\end{center}

It is shown that an alternative supersymmetric version
of the Liouville equation extracted from D=3 Green--Schwarz 
superstring equations \cite{lio} naturally arises as a super--Toda model 
obtained from a properly constrained  supersymmetric WZNW theory based on 
the $sl(2,{\bf R})$ algebra. Hamiltonian reduction is 
performed by imposing a nonlinear superfield constraint which turns out 
to be a mixture of a first-- and second--class constraint on 
supercurrent components. Supersymmetry of the model is realized 
nonlinearly and is spontaneously broken.\par
The set of independent current fields which survive the Hamiltonian reduction 
contains (in the holomorphic sector) one bosonic current of 
spin 2 (the stress--tensor of the spin 0 Liouville mode) 
and two fermionic fields of spin ${3\over 2}$ and $-{1\over 2}$.
The $n=1$ superconformal system thus obtained is of the same kind as one 
describing noncritical fermionic strings in a universal string theory 
\cite{ber}.
 
The generalization of this procedure allows one to produce 
from any bosonic Lie algebra super--Toda models
and associated super--${\cal W}$ algebras together with their 
nonstandard realizations.
\newpage
\renewcommand{\thefootnote}{\arabic{footnote}}
\setcounter{footnote}0
\section{Introduction}
In a recent paper \cite{lio} a system of constraints and equations of 
motion describing an $N=2$, $D=3$ Green--Schwarz superstring in a doubly 
supersymmetric (so called twistor--like) 
formulation \cite{gs,bpstv,bsv} was 
reduced to a superfield system of 
equations in $n=(1,1)$, $d=2$ worldsheet superspace 
\footnote{We denote the number
of left and right supersymmetries on the worldsheet by small $n$, 
and by capital $N$ the number of supersymmetries in target space}, 
which, in turn, was reduced to the purely 
bosonic Liouville equation plus two 
free massless fermion equations. 
Thus, the supersymmetric Liouville equation 
was shown to admit another $n=(1,1)$ supersymmetric realization  
different from the conventional one \cite{kulish,pol}.

In the present article we show that the same system of equations 
arises also in a properly constrained $n=(1,1)$ superconformal 
Wess--Zumino--Novikov--Witten model \cite{wznw} 
with an $Sl(2,{\bf R})$ group manifold as the target space.

Supersymmetric WZNW models based on classical Lie groups 
were considered in detail in \cite{v,a}.
A coset construction of superconformal field theories 
corresponding to constrained versions of the supersymmetric WZNW 
models was proposed in \cite{gko,coset} (for
a review see also \cite{halpern} and references therein). 
However the generalization of the Hamiltonian 
reduction \cite{bal,f} to supersymmetric WZNW models 
which leads to (super--) Toda 
models has been considered only for a restricted class of models, 
namely for those based on superalgebras which admit description in terms 
of fermionic simple roots only \cite{i,drs}. 
Such Hamiltonian reduction has 
not been applied to supersymmetric WZNW models 
based on bosonic groups and 
supergroups which have bosonic simple roots in any root decomposition
probably because of the lack of 
group--theoretical and physical grounds for imposing appropriate 
constraints on components of the WZNW currents, and also because of 
a simple argument that the Hamiltonian reduction 
should spoil supersymmetry of these models \cite{hollo}.

In the present paper we shall show how one can overcome this problem.
The model considered below is an 
example of the relation between strings propagating in flat space--time 
and appropriately constrained WZNW models, while
usually WZNW models are associated with strings propagating on group 
manifolds (see, for example, \cite{group}), or coset spaces (see, for 
example, \cite{costr}).  The WZNW group 
manifold under consideration is 
isomorphic to the structure group of the flat target superspace of 
the Green--Schwarz superstring. For $N=2$, $D=3$ target superspace of 
the Green--Schwarz superstring (as well as for $D=3$ 
fermionic strings)  
this group is $Sl(2,{\bf R})$. 
This provides us with a physical motivation to study the Hamiltonian 
reduction of wider class of the supersymmetric WZNW models, of which we 
shall consider in detail the simplest example, namely, 
the $n=(1,1)$, $d=2$ 
superconformal $Sl(2,{\bf R})$ WZNW model. 
An unusual feature we shall encounter with is that the nonlinear
superfield constraints imposed for performing the Hamiltonian reduction 
are a mixture of first-- and second--class constraints.
Supersymmetry of the model is realized nonlinearly and is 
spontaneously broken.
The set of independent currents which survive the Hamiltonian reduction 
contains (in the holomorphic sector) one bosonic current of 
spin 2 (which is the stress--tensor of the spin 0 Liouville mode) 
and two fermionic fields of spin ${3\over 2}$ and $-{1\over 2}$.
The $n=1$ superconformal system thus obtained is of the same kind as one 
used for describing a hierarchy of bosonic and 
fermionic strings embedded one into 
another \cite{ber}.
This demonstrates from a somewhat 
different point a (classical) relation between 
the Green--Schwarz and Neveu--Schwarz--Ramond--type formulation of 
superstrings \cite{gsw,vz,berk,berko}.

The prescription to be used can be 
generalized to obtain
super-Toda theories associated to any bosonic superaffine Lie algebra
and for constructing corresponding 
supersymmetric extensions of ordinary ${\cal W}$--algebras together with 
their non--standard realizations. 
The study of these models
may turn out to be useful, for instance, for getting new information 
about the structure of systems with spontaneously broken supersymmetry, 
their connection with the corresponding systems with linearly 
realized supersymmetry, 
and to gain deeper insight into the relationship 
between different superstring models.  
Some new structures can arise as a result of the Hamiltonian reduction of 
WZNW models based on superalgebras which contain simple bosonic roots.

The paper is organized as follows. In Section 2 we perform the 
Hamiltonian reduction of the supersymmetric $Sl(2,{\bf R})$ WZNW model 
and in Section 3 we demonstrate that the resulting system of equations 
is equivalent to that obtained in the $N=2$, $D=3$ Green--Schwarz 
superstring model \cite{lio}.
Symmetry properties and field contents 
of the system are analysed in Section 4. In 
Section 5 we make the Hamiltonian analysis of the constraints and get
(upon gauge fixing and constructing Dirac brackets) a Virasoro stress 
tensor and a supersymmetry current of an $n=1$ superconformal theory which 
describes an $n=(1,1)$, $D=3$ fermionic string in a physical gauge where 
two longitudinal bosonic degrees of freedom of the string are gauge 
fixed.   
In Section 6 we discuss
the general case of the Hamiltonian  
reduction of super---WZNW models based on the classical Lie groups to 
nonstandard supersymmetric versions of the Toda models. In Conclusion
we discuss an outlook of applying the procedure to study more general 
class of models.

\section{Supersymmetric $Sl(2,{\bf R})$ WZNW model and its reduction}
The superfield action of the model in $n=(1,1)$, $d=2$ superspace 
(with flat Minkowski metric)
parametrized by supercoordinates $Z=(z,\th)$, $\bar Z=(\bar z, \bar\th)$
(where $z,~\bar z$ are bosonic and $\th,~\bar\th$ are fermionic 
light--cone coordinates) has the form \cite{v,a}
\begin{equation}\label{1}
S = {k\over 2}\int dzd\bar z d\th d\bar\th\left(tr(G^{-1}DGG^{-1}\bar 
DG)+\int dt~ tr(G^{-1}{\partial\over{\partial t}}G
\{G^{-1}DG,G^{-1}\bar DG\})\right), 
\end{equation}
where $G(Z,\bar Z)$ is a superfield taking its values on the 
$Sl(2,{\bf R})$ group 
manifold, $D={\partial\over{\partial\th}}+
i\th{\partial\over{\partial z}}$ and 
$\bar D={\partial\over{\partial\bar\th}}+i\bar\th
{\partial\over{\partial\bar z}}$ are supercovariant fermionic 
derivatives which obey the following anticommutation relations
\begin{equation}\label{D}
\{D,D\}=2i{\partial\over{\partial z}},\qquad 
\{\bar D,\bar D\}=2i{\partial\over{\partial\bar z}}, \qquad
\{D,\bar D\}=0,
\end{equation}
and $k$ is called the level of the WZNW model (for simplicity in what 
follows we put $k=1$).
The second (Wess--Zumino) term in \p{1} is an integral over $d=3$,
$n=1$ superspace whose boundary is the ($Z,\bar Z$) superspace.

The action \p{1} is invariant under superconformal transformations 
of the $n=(1,1)$, $d=2$ superspace 
$$
Z'=Z'(Z),~~~~~{\bar Z}'={\bar 
Z}'(\bar Z);
$$ 
\begin{equation}\label{sc}
D'=e^{-\L}D,\qquad {\bar D}'=e^{-\bar\L}\bar D, \qquad 
{\rm where}~~~~\L=\log D\th'(Z),~~~~\bar\L=\log\bar D{\bar\th}'(\bar Z),
\end{equation}
and under the superaffine $Sl(2,\bf{R})$ transformations of 
$G(Z,\bar Z)$
\begin{equation}\label{2}
G'=g^{-1}_L(Z)G(Z,\bar Z)g_R(\bar Z).
\end{equation}

From \p{1} one gets the equations of motion
\begin{equation}\label{3}
\bar D(DGG^{-1})=0,\qquad D(G^{-1}\bar DG)=0,
\end{equation}
which read that the fermionic currents

\begin{eqnarray}\label{4}
\Psi(Z)\equiv {1\over i}DGG^{-1} &=\psi(z)+\th k(z)~~~~~~~~~~~~~~~~~\nn
&=\Psi^{-}E_-+\Psi^{+}E_{+}+\Psi^{0}H,~~ \nn
\bar\Psi(\bar Z)\equiv iG^{-1}\bar DG
&=\bar\psi(\bar z)+\bar\th \bar k(\bar z)~~~~~~~~~~~~~~~~~~\nn 
&=\bar\Psi^{-}E_-+\bar\Psi^{+}E_{+}+\bar\Psi^{0}H~~  
\end{eqnarray}
are (anti)chiral and hence conserved, and generate the superaffine 
$Sl(2,{\bf R})$ transformations \p{2}. $E_{\pm},~H$ are the 
$sl(2,{\bf R})$ algebra 
generators 
\begin{equation}\label{5}
[E_{+},E_{-}]=H,\qquad [H,E_{\pm}]=\pm 2E_{\pm}.
\end{equation}

The supersymmetric WZNW model can be constrained by putting equal 
to zero the 
components of the currents \p{4} corresponding to the generators 
\p{5} of a vector subgroup of the $Sl(2,{\bf R})_L\times Sl(2,{\bf 
R})_R$ affine group. On this way one gets coset construction of 
superconformal field theories \cite{gko,coset}.
Such constraints can be obtained from a 
gauged WZNW action \cite{gwz}, which 
has been under intensive study as an effective action for string 
configurations in target coset spaces (see, for example, \cite{costr}
and references therein).

Another possibility of reducing WZNW models (studied below)
is to impose constraints on those algebra--valued 
components of the WZNW currents which correspond to the nilpotent 
subalgebra of the positive (or negative) roots of the WZNW group 
generators \cite{bal,drs}. 
As we have already mentioned, this reduction establishes the 
relationship between (super) WZNW models \cite{bal,drs}
and (super) Toda theories \cite{ls}. 

In the case under consideration one cannot perform such kind of 
Hamiltonian reduction in a straightforward way. Indeed,  the 
basic supercurrents of the model are fermionic
\footnote{ It deserves  
mentioning that the (super)algebra--valued supercurrents of a
super--WZNW model have opposite statistic with respect to the
corresponding (super)algebra generators.},
and the Hamiltonian reduction procedure prescribes putting equal to 
nonzero constants those (super)currents which correspond 
to the positive or negative simple roots of 
the WZNW (super)algebra (a reduction of 
this kind leads to  so--called Abelian
Toda models, see \cite{bal,drs}).

If a current component to be constrained is fermionic, it does not  
seem natural to put it equal to a Grassmann constant. 
A way to overcome this problem is to explicitly use in the constraints
the Grassmann coordinates $\th$ or/and $\bar\th$ of superspace,
however the resulting model is no longer supersymmetric \cite{hollo}.
  
Therefore we shall use an alternative possibility, 
which does not violate manifest supersymmetry.
Firstly, we shall construct bosonic supercurrents 
out of fermionic ones \p{4} 
in an appropriate covariant way, and then impose constraints on 
the bosonic supercurrents. The natural candidates to be constrained are 
$sl(2)$--valued components of the bosonic supercurrents
\begin{eqnarray}\label{6}
J\equiv\partial GG^{-1}&=J^{+}E_{+}+J^{-}E_{-}+J^{0}H \nn
\bar J\equiv -G^{-1}\bar\partial G&
=\bar J^{+}E_{+}+\bar J^{-}E_{-}+\bar J^{0}H.
\end{eqnarray}
Together with the fermionic currents \p{4} the bosonic currents \p{6} 
constitute a differential one--superform being a pullback onto the 
$n=(1,1)$, $d=2$ superspace of the $Sl(2,{\bf R})$ Cartan form
\begin{equation}\label{7}
\Om=dGG^{-1},
\end{equation}
where 
$$
d=(dz-id\th\th){\partial\over{\partial z}}+
(d\bar z-id\bar\th\bar\th){\partial\over{\partial\bar z}}+d\th 
D+d\bar\th\bar D
$$
is the external differential.\footnote{Notice that, for instance, 
the ${d\bar\th}$ component of $\Omega$ is $\Om_{\bar D}={1\over 
i}G\bar\Psi G^{-1}$} By construction Eq. \p{7} satisfies the 
Maurer--Cartan equation (which is the zero curvature condition for the 
$Sl(2,{\bf R})$ connection form $\Om$)
\begin{equation}\label{8}
d\Omega-\Om\wedge\Om=0.
\end{equation}
From \p{8} it follows that the bosonic currents \p{6} are composed out 
of the fermionic currents \p{4} as follows
$$
J=D\Psi-i\Psi\Psi 
=k(z)-i\psi(z)\psi(z)+i\th(\partial\psi+\psi k-k\psi)
\equiv j(z)+i\th\chi(z),
$$
\begin{equation}\label{9}
\bar J=\bar D\bar\Psi -i\bar\Psi\bar\Psi
=\bar k(\bar z)-i\bar\psi(\bar z)\psi(\bar z)
+i\bar\th(\bar\partial\bar\psi+\bar\psi \bar k-\bar k\bar\psi)
\equiv \bar j(\bar z)+i\bar\th\bar\chi(\bar z).
\end{equation}

We are now in a position to constrain
the supersymmetric $Sl(2,{\bf R})$ 
WZNW model the same way as was used to reduce 
the bosonic $Sl(2,{\bf R})$ WZNW model to the 
Liouville theory \cite{{bal},{f}}. 
We put equal to one the current components 
$J^{-}$ and $\bar J^{+}$ of Eqs. \p{6}, \p{9}:
\begin{equation}\label{10}
J^{-}=D\Psi^{-}+2i\Psi^0\Psi^-=1, \qquad
\bar J^{+}=\bar D\bar\Psi^{+}-2i\bar\Psi^0\bar\Psi^+=1,
\end{equation}
where $D+2i\Psi^0$ and $\bar D-2i\bar\Psi^0$ can be regarded as 
covariant derivatives with $\Psi^0$, $\bar\Psi^0$ being components of 
a connection form on $d=2$ superspace. 
Eqs. \p{10} imply {\it nonlinear} relations between components 
of the fermionic currents \p{4} \footnote{n=(2,2) superconformal WZNW 
models and current algebras with nonlinear constraints on  
fermionic currents, which formally have the same form as the r.h.s. 
of Eq. \p{9}, were 
studied in \cite{h,ais}. Though the group--theoretical motivation for 
imposing the constraints in this form is the same as in our case, the 
constraints of \cite{h,ais} bear an essentially different 
meaning and result in 
unconstrained models in terms of n=(1,1) superfields \cite{h,ais}.}.

By using the Gauss decomposition of the $Sl(2,{\bf R})$ group element
\begin{equation}\label{G}
G(Z,\bar Z)=e^{\b E_+}e^{\Phi H}e^{\g E_-}
\end{equation}
one can (locally) write down the current components \p{4}
in the following form
$$
\Psi^-={1\over i}D\g e^{-2\Phi},\qquad
\Psi^+={1\over i}D\b-{1\over i}2\b D\Phi-\b^2\Psi^-,\qquad
\Psi^0={1\over i}D\Phi+\b\Psi^-;
$$
\begin{equation}\label{11}
\bar\Psi^+=i\bar D\b e^{-2\Phi},\qquad
\bar\Psi^-=i\bar D\g-2i\g \bar D\Phi-\g^2\bar\Psi^+,\qquad
\bar\Psi^0=i\bar D\Phi+\g\bar\Psi^+;
\end{equation}
Then the super--WZNW equations of motion \p{3} and the 
constraints \p{10} reduce to 
the following system of equations:
\begin{equation}\label{12}
\bar DD\Phi=e^{2\Phi}\bar\Psi^+\Psi^-, \qquad \bar 
D\Psi^-=0=D\bar\Psi^+;
\end{equation}
\begin{equation}\label{13}
D\Psi^{-}+2D\Phi\Psi^-=1, \qquad
\bar D\bar\Psi^{+}+2\bar D\Phi\bar\Psi^+=1.
\end{equation}
Eqs. \p{12} correspond to the unconstrained supersymmetric WZNW model.

The chirality conditions for the $\Psi^+$ and $\bar\Psi^-$ component of 
\p{11} are identically satisfied provided that \p{12} are valid. Note also 
that neither $\b(Z,\bar Z)$ nor $\g(Z,\bar Z)$ enters the Eqs. 
\p{12}, \p{13} explicitly if $\Psi^-$ and $\bar\Psi^+$ are considered as  
independent variables.

\section{Connection with N=2, D=3 Green--Schwarz superstrings}
In \cite{lio} the system of equations \p{12}, \p{13} was obtained from 
the equations of motions and constraints describing 
a classical $N=2$, $D=3$ 
Green-Schwarz superstring in a doubly supersymmetric 
geometrical approach \cite{gs,bpstv,bsv}.
To demonstrate this point we have to make a digression on ideas of 
applying the geometrical approach to describing the dynamics of 
(super)--p--branes.

The geometrical approach implies the use of notions and methods of 
surface theory \cite{eis} to describe embedding of p--brane 
worldvolumes into target spaces for a purpose of solving for the 
constraints and reducing the p--brane equations of motion to nonlinear
(sometimes exactly solvable) systems of equations for independent 
physical degrees of freedom of the p--branes. Its application to bosonic 
extended objects was initiated by Refs. \cite{lr,o}, and the 
reader may find a review of recent results in \cite{bn,las,bs} and 
references therein.

The development of supersurface theory to treat supergravities as 
theories of embedding supersurfaces into target supermanifolds was 
carried out in \cite{rs}.

A generalization of the geometrical approach to study super--p--branes 
stems from a doubly supersymmetric (so called twistor--like) formulation 
of these systems 
which is based on following principles (see \cite{bpstv} for details 
and references):

\noindent
i) the fermionic $\kappa$--symmetry of the Green--Schwarz formulation of
super--p--branes \cite{l,gsw} is a manifestation of 
superdiffeomorphisms of worldvolume
supersurfaces of the super--p--branes \cite{stv}. 
(This solves the problem of infinite
reducibility of the $\kappa$--symmetry and makes it possible to carry 
out covariant  Hamiltonian analysis of superstring dynamics at least on 
the classical level);

\noindent
ii) the geometrical ground for this is that the theory of super--p--branes is
supposed to be a particular kind of doubly supersymmetric models (studied
earlier in \cite{doubly}) which describe an embedding of supersurfaces into
target superspaces. (This naturally incorporates twistor--like commuting 
spinors into the theory as superpartners of target superspace Grassmann 
coordinates).

One of the aims of the approach has been to push forward the
problem of covariant quantization of superstring theory. As a result the
methods to attain this objective have undergone substantial modifications
and essential progress has been made
during last few years (see \cite{berko} for a recent review).

At the same time the doubly supersymmetric formulation 
serves as a natural dynamical basis for generalizing 
methods of classical surface theory to
study the embedding of supersurfaces
corresponding to super--p--branes \cite{bpstv}, and then one may try 
to apply the geometrical methods back to the analysis of variety 
of fundamental and solitonic super--p--branes we are having at hand at 
present time \cite{howes}. 

General properties of superstring and supermembrane 
worldvolumes embedded into flat target superspaces of various dimensions
were studied in \cite{bpstv} (see also \cite{band} for the case of 
N=1 superstrings). Using these results (which supergeneralize that of 
\cite{o})
we demonstrate below how by specifying the embedding of
worldsheet supersurface swept by the $N=2$, $D=3$ 
superstring
one can solve for the Virasoro constraints and reduce 
superstring equations of motion to the 
super--Liouville--like system of equations \p{12}, \p{13} 
\cite{lio}. Note that the classical equivalence of 
the twistor--like and Green--Schwarz formulation of the 
N=2, D=3 superstring was shown in \cite{gs}.

In the doubly supersymmetric formulation worldsheet of the Green--Schwarz
superstrings is a supersurface parametrized by two bosonic coordinates
$z=(\tau +\sigma)$, $\bar z=(\tau - \sigma)$  
and fermionic coordinates $\theta,~\bar\theta$ whose
number should be equal to the number of independent $\kappa$--symmetry
transformations in the standard Green--Schwarz formulation. In our case there
are one left-- and one right--handed Majorana--Weyl spinor coordinate 
$\theta, \bar\theta$, which
means that we deal with $n=(1,1)$ local supersymmetry on worldsheet
supersurface
\begin{equation}\label{ws}
{\cal M}_{ws}:~~~\left (Z=(z,\theta),~~~\bar Z=(\bar z, \bar 
\theta)\right).
\end{equation}

To describe intrinsic ${\cal M}_{ws}$ geometry 
(i.e. n=(1,1), d=2 supergravity)
one should set on ${\cal M}_{ws}$
a local frame of supervielbein one--forms which contains two bosonic vector
and to fermionic spinor components
\begin{equation}\label{e}
e^A(Z,\bar Z)=\left(e^a(Z,\bar Z), ~e^\a(Z,\bar Z)\right),
\end{equation}
where $a=(++,--)$ and $\a=(+,-)$ stand for light--cone vector and spinor 
indices of the $d=2$ Lorentz group $SO(1,1)$, respectively. 

We consider an embedding of ${\cal M}_{ws}$ into  $N=2$, $D=3$ flat
superspace--time paramet\-rized 
by three bosonic vector and two Majorana spinor
coordinates
$X^{\underline m}(Z,\bar Z),~\Theta^{1\underline \mu}(Z,\bar Z),$
$\Theta^{2\underline \mu}(Z,\bar Z),$
where the underlined indices 
${\underline m}=0,1,2$ and ${\underline \mu}=1,2$ are
vector and spinor indices of the $D=3$ Lorentz group $SO(1,2)\sim 
Sl(2,{\bf R})$, respectively.

A natural supersymmetric rigid frame in flat target superspace is 
\begin{equation}\label{rf}
\Pi^{\underline m}=dX^{\underline m}+id\bar\Theta^{i}
\Gamma^{\underline m}\Theta^{i}, \qquad d\Theta^{i\underline \mu} \qquad
(i=1,2).
\end{equation}
$\Gamma^{\underline m}_{\underline{\a\b}}$ are $D=3$ Dirac matrices.

The study of ${\cal M}_{ws}$ embedding is started with fitting the rigid
target--superspace frame \p{rf} to that on the supersurface \p{e}.
To this end we transform \p{rf} into a new local frame 
\begin{equation}\label{lf}
E^{\underline a}=\Pi^{\underline m}
u_{\underline m}^{\underline a}(X,\Theta),\qquad
E^{i \underline \alpha}=d\Theta^{i\underline \mu}v_{\underline \mu}^
{\underline \alpha}(X,\Theta),
\end{equation}
by use of $u_{\underline m}^{\underline a}$ and $v_{\underline\mu}^
{\underline \alpha}$ matrices of the vector and spinor
representation of the target--superspace Lorentz group $Sl(2,{\bf R})$, 
respectively. But since the vector and spinor
components of \p{rf} are subject to the Lorentz transformation simultaneously,
$u$ and $v$ are not independent and connected through 
the well--known twistor--like relation
$$
u_{\underline m}^{\underline a}(\Gamma^{\underline m})_{\underline
{\mu\nu}}= v^{\underline \alpha}_{\underline \mu}
(\Gamma^{\underline a})_{\underline{\alpha\beta}}
v^{\underline \beta}_{\underline \nu}
$$
between vectors and commuting spinors.
This explains why the approach is called ``twistor--like".

From the general analysis of superstring dynamics in the 
twistor--like approach we learn \cite{bpstv} 
that the target--superspace local frame \p{lf} can be attached
to the supersurface as follows:
\begin{equation}\label{pull}
E^{\perp}(Z,\bar Z) = 0, 
\end{equation}
\begin{equation}\label{pulla}
E^a(Z,\bar Z) = \Pi^{\underline m}u^a_{\underline m}=
(dX^{\underline m}+id\bar\Theta^{i}
\Gamma^{\underline m}\Theta^{i})u^a_{\underline m}=e^a,
\end{equation}
\begin{equation}\label{pullb}
E^{1+}(Z,\bar Z) = d\Theta^{1\underline \mu}v^{+}_{\underline \mu}=e^+,
\qquad
E^{2-}(Z,\bar Z) = d\Theta^{2\underline \mu}v^{-}_{\underline \mu}=e^-,
\end{equation}
where the target superspace indices split onto that of ${\cal M}_{ws}$ 
and of the orthogonal vector direction 
(${\underline a}\rightarrow (a,\perp);~ {\underline \alpha}\rightarrow 
(+,-))$.

Eq. \p{pull} tells us that one of the vector components of
the target superspace frame can be made orthogonal to the supersurface
(its pullback on ${\cal M}_{ws}$ is zero) and three other
relations \p{pulla}, \p{pullb} identify (on ${\cal M}_{ws}$) 
components of the target superspace
frame with the intrinsic ${\cal M}_{ws}$ supervielbein components \p{e}.
This means that ${\cal M}_{ws}$ geometry is induced by embedding.

From \p{pull} and \p{pulla}, using the orthogonality properties
of $u_{\underline m}^{\underline a}$ ($u_{\underline m}^{\underline a}
u_{\underline n \underline a}=\eta_{{\underline m \underline n}}$), 
we find that the pullback on ${\cal M}_{ws}$  of the vector one--superform
$
\Pi^{\underline m}(Z,\bar Z)=dX^{\underline m}+id\bar\Theta^{i}
\Gamma^{\underline m}\Theta^{i}
=e^au_a^{\underline m}(Z,\bar Z)
$
is zero along the
fermionic directions $e^{\a}$ of ${\cal M}_{ws}$:
\begin{equation}\label{gdc}
DX^{\underline m}+iD\bar\Theta^{i}
\Gamma^{\underline m}\Theta^{i}=0,
\qquad
\bar DX^{\underline m}+i\bar D\bar\Theta^{i}
\Gamma^{\underline m}\Theta^{i}=0,
\end{equation}
where $D,\bar D$ are covariant spinor derivatives on ${\cal M}_{ws}$ 
which reduce to the flat derivatives \p{D} in a superconformal gauge for 
\p{e}.
Eq. \p{gdc} is called the geometrodynamical condition. 
Eqs. \p{pull} and \p{pulla} ensure that the
Virasoro constraints on superstring dynamics
\begin{equation}\label{vc}
\Pi^{\underline m}_{++}\Pi^{\underline m}_{++}=0=
\Pi^{\underline m}_{--}\Pi^{\underline m}_{--}
\end{equation}
are identically satisfied for such kind of embedding \cite{bpstv}. 

One can notice that Eqs. \p{pull}--\p{pullb} 
are first--order differential
equations on $X(Z,\bar Z)$ and $\Theta(Z, \bar Z)$. 
If they are solved one would know the
shape of the worldsheet supersurface in $D=3$, $N=2$ superspace and thus 
would
solve the problem of describing classical superstring motion.
To solve \p{pull}--\p{pullb} 
one must know $v^{\underline \alpha}_{\underline \mu}(Z,\bar Z)$
and the components of $e^a(Z,\bar Z), e^{\a}(Z,\bar Z)$. 
To get this information one should study the integrability conditions of 
\p{pull}--\p{pullb} 
which are obtained by taking the external differential
of \p{pull}--\p{pullb}. (This is the general strategy one pursues in 
the geometrical approach \cite{lr}--\cite{bn}, \cite{bpstv}). 
Basic integrability conditions thus obtained are \cite{bpstv,bsv}:
\begin{equation}\label{cond}
de^a-\Omega^a_b e^b=ie^\a\gamma^a_{\a\b}e^{\b}=T^a,
\end{equation}
\begin{equation}\label{sec}
\Om^{\perp a}=K^a_b e^b+K^a_\a e^\a,
\end{equation}
where $\gamma^a_{\a\b}$ are d=2 Dirac matrices, and external
differentiation and product of the forms are implied.
Eqs. \p{cond}, \p{sec} contain one forms $\Omega^a_b, \Om^{\perp a}$ which are
Cartan forms of the $Sl(2,{\bf R})$ 
Lorentz group \p{7} constructed out of the matrix
$v^{\underline \alpha}_{\underline \mu}$ components:
\begin{equation}\label{cartan}
\Om^{ab}=\e^{ab}v^{+}_{\underline \mu}dv^{-\underline \mu},
\qquad
\Om^{\perp a}=\gamma^a_{\a\b}v^{\alpha}_{\underline \mu}
dv^{\b\underline \mu}.
\end{equation}

Eq. \p{cond} determines parallel transport of vector supervielbeins along
 ${\cal M}_{ws}$ carried out by induced connection $\Om^{ab}$. It reads
that the connection possesses torsion whose spinor--spinor components are
constrained to be equal to the $\g$--matrix components. This is a basic
torsion constraint of any supergravity theory. In the
geometrical approach it is not imposed by hand, but appears, together 
with other $n=(1,1)$, $d=2$ supergravity constraints \cite{d2}, as a
consistency condition of ${\cal M}_{ws}$ embedding \p{pull}--\p{pullb} 
\cite{bsv,bpstv}. From \cite{d2} we know that d=2 supergravity thus 
constrained is superconformally flat. This means that by gauge fixing 
${\cal M}_{ws}$ superdiffeomorphisms one can choose a superconformal 
gauge where the supervielbeins \p{e} take the form
$$
e^{++}=e^{-2(\Phi-L)}(dz-id\th\th),\qquad
e^{--}=e^{-2(\Phi+L)}(d\bar z-id\bar\th\bar\th),
$$
\begin{equation}\label{suc}
e^+=e^{-(\Phi-L)}\left(d\th+iD\Phi(dz-id\th\th)\right),\qquad
e^-=e^{-(\Phi+L)}\left(d\bar\th+
i\bar D\Phi(d\bar z-id\bar\th\bar\th)\right),
\end{equation}
where the superfield $\Phi(Z,\bar Z)$ corresponds to Weyl rescaling and
$L(Z,\bar Z)$ corresponds to local Lorentz $SO(1,1)$
transformations of the supervielbeins. Note that Weyl transformations 
are not a symmetry of the model, while the local $SO(1,1)$ 
transformations are \cite{bpstv}.

The second condition \p{sec} specifies the expansion of $\Om^{\perp a}$ in
the ${\cal M}_{ws}$ supervielbein components: the bosonic matrix
$K_{ab}(Z)$ is symmetric and has the properties of the second fundamental
form analogous to that of the bosonic surfaces, while spinor components
$K_\a(Z)\equiv K^a_{\b}\gamma_{a\a}^{\b}$ of the Grassmann--odd spin--tensor
$K^a_{\b}$ can be associated with a fermionic counterpart of the second
fundamental form along Grassmann directions of the supersurface 
\cite{bpstv}.

By construction the Cartan forms \p{cartan} must satisfy the 
$Sl(2,{\bf R})$
Maurer--Cartan equations \p{8} which split into two systems of
equations with respect to the worldsheet indices:
\begin{equation}\label{cod}
d\Om^{\perp a}-\Om^a_b\Om^{\perp b}=0,
\end{equation}
\begin{equation}\label{gauss}
R^{ab}=d\Om^{ab}=\Om^{\perp a}\Om^{\perp b}.
\end{equation}
Eq. \p{cod} is known as the Codazzi 
equation and \p{gauss} is called the Gauss
equation in surface theory. On the other hand one can recognize in \p{cod}
and \p{gauss} relations which specify geometry on a two--dimensional coset
space ${SO(1,2)}\over{SO(1,1)}$ with $\Om^{\perp a}$ being a vielbein,
$\Om^{ab}$ being a spin connection and $R^{ab}$ being a constant curvature
tensor of  ${SO(1,2)}\over{SO(1,1)}$.

Thus we have reduced the problem of studying superstring dynamics (as
the embedding of a supersurface into target superspace) 
to study a mapping
of ${\cal M}_{ws}$ onto the bosonic coset space of constant curvature.
It is here that a connection of Green--Schwarz superstring dynamics with 
the $n=(1,1)$ super--WZNW model comes out.

To completely describe superstring dynamics in geometrical terms we should
specify what additional conditions on embedding arise when the superstring
equations of motion are taken into account \cite{lr,o,bn,bpstv}. 
In bosonic surface theory
such an embedding is called minimal 
and is characterized by traceless second
fundamental form $K_{ab}$.
In the supersymmetric case we get analogous condition on the bosonic part of
the second fundamental form \p{sec} which is in one to one correspondence
with $X^{\underline m}$ equations of motion
\begin{equation}\label{trl}
K^a_a\equiv
D_{a}\left(D^{a}X^{\underline m}-i\bar\Theta^{i}
\Gamma^{\underline m}D^{a}\Theta^{i}\right)u_{\underline m}^{\perp}=0,
\end{equation}
where $D_a=(D_{--},D_{++})$ is a vector covariant derivative, which 
reduces to (${\partial\over{\partial z}},{\partial\over{\partial\bar 
z}}$) in the superconformal gauge. In
addition, $\Theta(Z, \bar Z)$ equations of motion, which in the twistor--like
approach have the form
$$
K_-\equiv D_{--}\Th^{1\underline \mu}v^-_{\underline \mu}=0,
\qquad
K_+\equiv D_{++}\Th^{2\underline \mu}v^+_{\underline \mu}=0,
$$
result in vanishing the fermionic counterpart 
of the second fundamental form in Eq. \p{sec}, namely
\begin{equation}\label{f}
K_\a(Z,\bar Z)\equiv K^a_{\b}\gamma_{a\a}^\b=0, \qquad \a =(+,-).
\end{equation}

Thus we see that in the geometrical approach the dynamical string 
equations of motion are replaced by algebraic conditions (constraints)
on components of the Cartan form
\p{sec}, and the role of the dynamical equations is taken by the 
Maurer--Cartan equations \p{cod}, \p{gauss} (which reminds the reduction 
of the WZNW model).

The minimal embedding conditions \p{trl}, \p{f} further reduce the number
of independent superfields which determine the induced geometry on
the worldsheet supersurface ${\cal M}_{ws}$. 
One can show \cite{lio,band} that in the superconformal gauge \p{suc}
$\Om^{\perp a}$ and $\Om^{ab}$ 
(which now, by virtue of theorems of surface 
theory, bear all information about
${\cal M}_{ws}$) depend only on two bosonic 
superfields $L(Z,\bar Z)$ and
$\Phi(Z,\bar Z)$ \p{suc}, 
and two fermionic superfields $\bar\Psi^+(\bar Z)$ and 
$\Psi^-(Z)$.
The leading components of $\Phi(Z,\bar Z)$, $\bar\Psi^+(\bar Z)$ and 
$\Psi^-(Z)$ describe one bosonic and two fermionic 
physical degrees of freedom of the classical N=2, D=3 Green--Schwarz
superstring. As to the superfield $L(Z,\bar Z)$, we have already
mentioned that it corresponds to purely 
gauge degrees of freedom reflecting local worldsheet $SO(1,1)$ 
invariance of the superstring model. The exact form of the basic 
fermionic components of the Cartan forms \p{cartan} is \cite{lio}
(in our convention)
$$
F^-\equiv {1\over i}v^{-}_{\underline \mu}
Dv^{-\underline \mu} =\Psi^-e^{\Phi-L},~~~
F^+\equiv {1\over i}v^{+}_{\underline \mu}
Dv^{+b\underline \mu}=0,~~~F^0\equiv {1\over i}v^{+}_{\underline \mu}
Dv^{+\underline \mu}={1\over{2i}}(D\Phi+DL);
$$
\begin{equation}\label{16}
\bar F^+\equiv iv^{+}_{\underline \mu}
\bar Dv^{+b\underline \mu}=\bar \Psi^+e^{\Phi+L},~~~
\bar F^-\equiv iv^{-}_{\underline \mu}
\bar Dv^{-\underline \mu}=0,~~~
\bar F^0\equiv iv^{+}_{\underline \mu}
\bar Dv^{+\underline \mu}={i\over{2}}(\bar D\Phi-\bar DL).
\end{equation}
The bosonic components of the Cartan forms \p{cartan} are 
not independent and expressed in terms of \p{16} (similar to \p{9}) 
through the Maurer--Cartan equations \p{8} (or \p{cod}, \p{gauss}),
and the superfields \p{16} themselves
obey the $n=(1,1)$ superconformal invariant equations 
\p{12}, \p{13} which now follow from \p{gauss} and \p{cod} \cite{lio}.
One should not confuse the Liouville system describing classical 
physical modes of the string with (super)Liouville modes emerging as 
anomalies of quantized noncritical strings \cite{pol}. 
This completes our sketch of how Eqs. \p{12}, \p{13} emerge in the 
$N=2$, $D=3$ superstring model.

Note that in the process of reducing the problem of describing classical 
$N=3$, $D=3$ superstring dynamics to the Codazzi--Gauss equations 
\p{cod}, \p{gauss} we have hidden $N=2$, $D=3$ space--time 
supersymmetry since the ${\cal M}_{ws}$ differential one--forms 
\p{cartan} take their values in the bosonic $sl(2,{\bf R})$ algebra and 
describe ${\cal M}_{ws}$ mapping onto the bosonic coset space. Thus in 
the geometrical approach Green--Schwarz  superstrings look much more 
like fermionic strings which {\it a priori} propagate in bosonic 
space--time. We shall turn back to this point in Section 5.

Let us now compare the $Sl(2,{\bf R})$ Cartan form components \p{16}
with the $Sl(2,{\bf R})$ WZNW fermionic currents of the previous 
section.
We see that the components of \p{16} do not have the form of 
the Gauss decomposition, they are nonchiral and, hence, differ from 
the currents \p{11}. However, $F,~\bar F$ and $\Psi,~\bar\Psi$ relate to 
each other the same way as a zero curvature connection of a bosonic Toda 
theory \cite{ls} relates to (anti)chiral currents of the corresponding 
reduced WZNW model \cite{f}. To establish the relationship 
between Eqs. \p{16} and Eqs. \p{11} one 
should perform in \p{16} an independent left and 
right gauge transformation of  
$
F= F^- E_- + F^0 H + F^+E_+ $ and 
$
{\bar F} = {\bar F}^- E_-
+ {\bar F}^0 H + {\bar F}^+ E_+$,
which correspond, respectively, to 
the $\beta(Z,\bar Z)$ and $\g(Z,\bar Z)$ 
factor of the Gauss decomposition 
\p{G} of an $Sl(2,{\bf R})$ group element:
\begin{equation}\label{17a}
\hat F={1\over i}D\b E_++e^{\b E_+}Fe^{-\b E_+}, \qquad
\hat{\bar F}={1\over i}\bar D\b E_++e^{\b E_+}\bar Fe^{-\b E_+}, 
\end{equation}
\begin{equation}\label{17b}
\hat F={i}D\g E_-+e^{-\g E_-}Fe^{\g E_-}, \qquad
\hat{\bar F}={i}\bar D\g E_-+e^{-\g E_-}\bar Fe^{\g E_-}.
\end{equation}
To perform an appropriate $\b$--transformation \p{17a} we first gauge fix 
$L=\Phi$ in \p{16}, and then choose $\b(Z,\bar Z)$ to satisfy the 
condition 
$$
\bar D\beta={1\over i}\bar\Psi^+e^{2\Phi}.
$$
Then the $\hat{\bar F}$ components in \p{17a} turn to zero and ${\hat 
F}^{\mp},~{\hat F}^0$ coincide with $\Psi^{\mp},~\Psi^0$, respectively.

On the  other hand, to perform an appropriate 
$\g$--transformation we choose 
in \p{16} $L=-\Phi$ and take $\gamma(Z,\bar Z)$ to satisfy 
$$
D\g={i}\Psi^-e^{2\Phi}.
$$
Then the $\hat F$ components of \p{17b} turn to zero and
$\hat{\bar F}^{\pm},~\hat{\bar F}^0$ 
reduce to $\bar\Psi^{\pm},~\bar\Psi^0$. Notice that the gauge 
transformation parameters $\beta(Z,\bar Z)$ and $\g(Z,\bar Z)$ 
indeed coincide with those of Eqs. \p{11}.

To conclude this section we should point to a problem which we have not 
solved. It is the problem of constructing a gauged version of the WZNW 
action \p{1} from which the constraints \p{10} could be obtained as the 
equations of motion of auxiliary gauge fields. The straightforward 
application of the procedure used for gauging the WZNW models subject
to `standard' Hamiltonian reduction \cite{bal,f,drs} 
does not work, since in our 
case the constraints \p{10} are nonlinear, contain supercovariant 
derivatives of fermionic currents and, as we shall see in the next 
section, are a mixture of first-- and second--class constraints.
However an indirect indication that such a gauged supersymmetric WZNW 
action may exist is that there exist versions of the $N=2$, $D=3$ 
superstring action with local $n=(1,1)$ worldsheet supersymmetry 
\cite{gs,bsv} 
from which one can get the system of equations \p{12}, \p{13}. As we have 
demonstrated, the currents that arise in the superstring model are 
connected with the WZNW currents by the local transformations \p{17a}, 
\p{17b}, which might be associated with local symmetry transformations of 
the gauged WZNW action if the latter existed.

\section{Superconformal properties of the model and of 
the related Liouville system}
Let us consider now superconformal properties of  
fields subject to the constraints \p{13}. The Eqs. \p{12}, \p{13} are 
invariant under the $n=(1,1)$ transformations \p{sc} provided that
the latter are accompanied by the following left--right $H$--rotation 
of $G(Z,\bar Z)$ \p{G}:
\begin{equation}\label{H}
\hat G=e^{-(\L+{1\over 2}D\L\Psi^-)H}G
e^{-(\bar\L+{1\over 2}\bar D\bar\L\bar\Psi^+)H}.
\end{equation}
Then, due to \p{11}, 
\begin{equation}\label{20}
\hat\Phi=\Phi-(\L+\bar\L)-{1\over 2}(D\L\Psi^-+\bar D\bar\L\bar\Psi^+),
\qquad \hat\b=e^{-2\L-D\L\Psi^-}\b, \qquad 
\hat\g=e^{-2\bar\L-\bar D\bar\L\bar\Psi^+}\g,
\end{equation}
and
\begin{equation}\label{21}
{\hat\Psi}^-=e^\L\Psi^-, \qquad 
{\hat\Psi}^+=e^{-3\L-D\L\Psi^-}\Psi^+, \qquad
{\hat\Psi}^0=e^{-\L}({\Psi}^0-{1\over i}D(\L+{1\over 2}D\L\Psi^-)),
\end{equation}
\begin{equation}\label{211}
\hat{\bar\Psi}^+=e^{\bar\L}\bar\Psi^+, \qquad
\hat{\bar\Psi}^{-}=e^{-3\bar\L-\bar D\bar\L\bar\Psi^+}\bar\Psi^-, \qquad
\hat{\bar\Psi}^{0}=e^{-\bar\L}({\bar\Psi}^0-{i}\bar D
(\bar\L+{1\over 2}\bar D\bar\L\bar\Psi^+)).
\end{equation}

From Eqs. \p{20} -- \p{211} we see that  the supercurrents $\Psi^+$ and 
$\bar\Psi^-$ transform nonlinearly under the modified $n=(1,1)$ 
superconformal transformations, and even
though the superfields $\Phi$, $\Psi^-$, $\bar\Psi^+$, $\Psi^0$ and 
$\bar\Psi^0$ transform linearly, 
the $n=(1,1)$ superconformal symmetry is nonlinearly realized on the 
components of the superfields, since they are constrained by Eqs.
\p{13} (the basic fields belong in fact to different supermultiplets). 

The general solution to \p{13} (obtained by taking into account  
(anti)chirality of $\Psi^-$, $\bar\Psi^+$) is (\cite{lio}):
\begin{equation}\label{22}
\Psi^-=\psi^-(z)+\th(1+i\psi^-\partial\psi^-), \qquad
\bar\Psi^+=\bar\psi^+(\bar z)+
\bar\th(1+i\bar\psi^+\bar\partial\bar\psi^+),
\end{equation}
\begin{equation}\label{23}
\Phi=\phi(z,\bar z)+{i\over 2}\th e^{-2\phi}\partial(e^{2\phi}\psi^-)+
{i\over 2}\bar\th e^{-2\phi}\bar\partial(e^{2\phi}\bar\psi^+)+
\th\bar\th\psi^-\bar\psi^+\partial\bar\partial\phi.
\end{equation}

From Eqs. \p{sc}, \p{20} -- \p{23} it follows that under infinitesimal
(anti)holomorphic supersymmetry transformations 
$\th\rightarrow\th-\e(z)$, $\bar\th\rightarrow\bar\th-\bar\e(\bar z)$
the leading component of $\Phi$ transforms as follows
\begin{equation}\label{24.}
\phi(z,\bar z)~\rightarrow~\phi(z,\bar z)+i\e(z)\psi^-\partial\phi
+{i\over 2}\partial(\e(z)\psi^-)
+i\bar\e(\bar z)\bar\psi^+\bar\partial\phi
+{i\over 2}\bar\partial(\bar\e(\bar z)\bar\psi^+),
\end{equation}
and
the leading components of $\Psi^-$, $\bar\Psi^+$ transform as Goldstone 
fermions \cite{av}
\begin{equation}\label{24}
\psi^-(z)~\rightarrow~\psi^-(z)+\e(z)+i\e(z)\psi^-\partial\psi^-,
\qquad
\bar\psi^+(\bar z)~\rightarrow~\bar\psi^+(\bar z)+\bar\e(\bar z)+
i\bar\e(\bar z)\bar\psi^+\partial\bar\psi^+,
\end{equation}
which signifies that the $n=(1,1)$ supersymmetry is spontaneously 
broken. 
 Note that the form of the transformations \p{24} does not imply 
that $\psi^-(z)$, $\bar\psi^+(\bar z)$ are pure gauge degrees of 
freedom, since here superconformal symmetry 
is not a local symmetry in a full 
sense and cannot be used for reducing the number of the physical degrees 
of freedom of the model (there is no first--class constraints which 
generate the superconformal symmetry).

The supersymmetry transformation properties of the leading components of
$\Psi^+$ and $\Psi^0$ ($\bar\Psi^-$ and $\bar\Psi^0$ in the 
antiholomorphic sector) are
$$
\d\psi^+=(j^++2i\psi^0\psi^+)\e+i\psi^-\psi^+\partial\e, \qquad
\d\psi^0=(j^0+i\psi^+\psi^-)\e+{1\over 2}(1+2i\psi^-\psi^0)\partial\e,
$$
\begin{equation}\label{+0}
\bar\d\bar\psi^-=(\bar j^-+2i\bar\psi^0\bar\psi^-)\bar\e+
i\bar\psi^+\bar\psi^-\partial{\bar\e}, \qquad
\bar\d\bar\psi^0=(\bar j^0+i\bar\psi^-\bar\psi^-)\bar\e+
{1\over 2}(1+2i\bar\psi^+\bar\psi^0)\partial{\bar\e}.
\end{equation}

Under conformal transformations 
($z~\to~z-\l(z)$, $\bar z~\to~\bar z-\bar\l(\bar z)$,) the fields
$\phi$, $\psi^-$, $\bar\psi^+$ transform as follows:
\begin{equation}\label{conb}
\d\phi=\l\partial\phi+{1\over 2}\partial\l+
\bar\l\bar\partial\bar\phi+{1\over 2}\bar\partial\bar\l,
\end{equation}
\begin{equation}\label{conf}
\d\psi^-=\l\partial\psi^- -{1\over 2}\psi^-\partial\l, \qquad
\bar\d\bar\psi^+=\bar\l\bar\partial\bar\psi^+ 
-{1\over 2}\bar\psi^+\bar\partial\bar\l,
\end{equation}

and
\begin{equation}\label{confo}
\d\psi^+=\l\partial\psi^+ +{3\over 2}\psi^-\partial\l, \qquad
\d\psi^0=\l\partial\psi^- +{1\over 2}\psi^0\partial\l +{1\over 
4}\psi^-\partial^2\l,
\end{equation}
\begin{equation}\label{confor}
\bar\d\bar\psi^-=\bar\l\bar\partial\bar\psi^- +
{3\over 2}\bar\psi^+\bar\partial\bar\l, \qquad
\bar\d\bar\psi^0=\bar\l\bar\partial\bar\psi^+ 
+{1\over 2}\bar\psi^0\bar\partial\bar\l +
{1\over 4}\bar\psi^+{\bar\partial}^2\bar\l.
\end{equation}

The transformations \p{24.} -- \p{confor}
form an $n=(1,1)$ superconformal algebra which
closes on the mass shell. For instance
\begin{equation}\label{cl}
\left[\d_{\e_1},\d_{\e_2}\right]=\d_\l,
\end{equation}
where $\l=2i\e_1\e_2$.
From \p{conf} we see that $\psi^-$, $\bar\psi^+$ have conformal spin 
$-{1\over 2}$, and $\psi^-$, $\bar\psi^+$ have spin ${3\over 2}$. 
As to the fields $\psi^0$, $\bar\psi^0$, their transformation 
properties \p{+0}, \p{confo}, \p{confor} reflect the fact that, due to 
the constraints \p{10}, these fields are not independent (see next 
Section for the details). 

Comparing \p{24.} with \p{conb} one can notice that these 
transformations have the same form for $\l=i\e\psi^-$ and 
$\bar\l=i\bar\e\bar\psi^+$. Thus, in some sense, supersymmetric 
transformations of $\phi$ can be hidden in its conformal 
transformations. This effect was observed by the authors of \cite{v}
in the unconstrained supersymmetric $O(N)$ WZNW model. Note also that 
the  supersymmetric
transformations of $\psi^-$, $\bar\psi^+$ 
contain only the fermionic fields themselves,
nevertheless, as we have seen (Eq. \p{cl}), the commutator of 
two supersymmetry transformations leads to a conformal transformation, 
therefore the superconformal symmetry of our model is fully fledged in 
contrast to a Grassmann symmetry of a free fermion system of Ref. 
\cite{v}.

Finally, we present the component equations resulting from \p{12}, 
\p{13} \cite{lio}:
\begin{equation}\label{l} 
\partial\bar\partial\phi=-e^{2\phi},
\end{equation}
\begin{equation}\label{c}
\bar\partial\psi^-=0,\qquad \partial\bar\psi^+=0.
\end{equation}
As one might expected, the reduction of the supersymmetric $Sl(2,{\bf R})$ 
WZNW model resulted in the Liouville equation \p{l}, which  
is accompanied by the free chiral fermion equations \p{c}.
An interesting point is that  
the whole system of the equations \p{l}, \p{c}
is invariant under the nonlinear 
$n=(1,1)$ superconformal transformations \p{24.}, \p{24}, and thus can 
be considered as a supersymmetric generalization of the Liouville 
equation alternative \cite{lio} to the conventional one based on the 
supergroup $OSp(1|2)$ \cite{kulish}. 
\footnote{In \cite{lio} there was an overstatement 
that the two super--Liouville 
systems are not connected with each other by local redefinition of 
fields. In fact, this is only true one way, i.e. when one tries to get  
the free fermion equations from the standard super--Liouville equations
\cite{kulish}. But it turns out possible to locally express the standard 
super--Liouville fields through the fields of \p{l}, \p{c} \cite{ikp}.}
Together with this system one should consider 
free fermionic fields $\psi^+(z)$, 
$\bar\psi^-(z)$ of spin ${3\over 2}$ which also remain independent after 
the Hamiltonian reduction. The properties of the whole system are 
studied in the next Section.

\section{Current algebra, Hamiltonian analysis of the constraints and 
connection with fermionic strings}
Poisson brackets of the fermionic superfield currents \p{4}  
generate two (anti)com\-mu\-ting copies of the superaffine $SL(2,{\bf 
R})$ algebra \cite{drs}
\begin{equation}\label{pbs}
\left\{tr(A\Psi(X)),tr(B\Psi(Y))\right\}_{PB}
=-\d(X-Y)tr[A,B]\Psi(Y)-iD_X\d(X-Y)tr(AB),
\end{equation}
where $A,~B$ stand for $E_+,~E_-$ and $H$; 
$X=(z_1,\th_1)$, $Y=(z_2,\th_2)$, \\
$\d(X-Y)=\d(z_1-z_2)(\th_1-\th_2)$, and
$D_X={\partial\over\partial\th_1}+i\th_1{\partial\over\partial z_1}$.\\
It is implied that the Poisson brackets are equal--time, i.e.
$z_1+\bar z_1=z_2+\bar z_2$. 
Since the Poisson brackets of $\bar\Psi(\bar Z)$ have the same form as 
\p{pbs}, we restrict ourselves to the consideration of only the holomorphic 
sector of the model.

From \p{pbs} one can get the Poisson brackets of the bosonic superfield 
currents $J(Z)$ \p{9}. We do not present them explicitly because of a 
somewhat cumbersome structure of their r.h.s. Instead we shall deal with 
Poisson brackets of the ordinary field components 
of $\Psi(Z)$ \p{4} and $J(Z)$
\p{9}. It is convenient to work with $\psi(z)$ and $j(z)$ as the 
independent components of the superfield currents, 
since they are completely decoupled \cite{v,a}:
\begin{eqnarray}\label{pbc}
\{j^+(z_1),j^-(z_2\}_{PB}& =&2\d(z_1-z_2)j^0(z_2)-\partial\d(z_1-z_2),\nn
\{j^0(z_1),j^{\pm}(z_2)\}_{PB}& = & \pm \d(z_1-z_2) j^{\pm}(z_2),\nn
\{j^0,j^0\}_{PB}& = & -{1\over 2}\partial\d(z_1-z_2) \nn
\{j^+,j^+\}_{PB}& = & \{j^-,j^-\}_{PB} =0,\\
\{j^a,\psi^b\}_{PB}& = & 0, \nn
\{\psi^+(z_1),\psi^-(z_2)\}_{PB}& = &- i\d(z_1-z_2),\nn
\{\psi^0(z_1),\psi^0(z_2)\}_{PB}& = & -{i\over 2}\d(z_1-z_2),\nn
\{\psi^+,\psi^+\}_{PB}& = & \{\psi^-,\psi^-\}_{PB} = 0.\nonumber
\end{eqnarray}
Using \p{pbc} it is rather easy to understand the structure of the 
superfield constraints \p{10}, which, in components, reduce to
\begin{equation}\label{con}
j^--1=0,\qquad \psi^0-j^0\psi^--{1\over 2}\partial\psi^-=0
\end{equation}
(in the holomorphic sector).

A linear combination of the constraints \p{con}
\begin{equation}\label{cf}
C_{\scriptscriptstyle F}=\psi^0-j^0\psi^--{1\over 2}\partial\psi^-
+{1\over 4}\partial\psi^-(j^--1)=0,
\end{equation}
\begin{equation}\label{cb}
C_{\scriptscriptstyle B}=j^-(1+2i\psi^-C_{\scriptscriptstyle F})-1=0
\end{equation}
have the following form of the Poisson brackets \p{pbc}:
\begin{equation}\label{1st}
\{C_{\scriptscriptstyle B},C_{\scriptscriptstyle B}\}_{PB}=
\{C_{\scriptscriptstyle B},C_{\scriptscriptstyle F}\}_{PB}=0,
\end{equation}
\begin{equation}\label{2nd}
\{C_{\scriptscriptstyle F}(z_1),C_{\scriptscriptstyle F}(z_2)\}_{PB}
=-{i\over 2}\d(z_1-z_2).
\end{equation}
Note that the equations \p{cf} -- \p{2nd} are satisfied in a weak sense, 
i.e. $C_{\scriptscriptstyle B}(Z)$ and $C_{\scriptscriptstyle F}(Z)$ can 
be put to zero only upon calculating 
their Poisson brackets with other field expressions. From \p{1st}, 
\p{2nd} we conclude that the bosonic constraint \p{cb} is of the first 
class and the fermionic constraint \p{cf} is of the second class. The 
latter can be put to zero in the strong sense by replacing the Poisson 
brackets with Dirac brackets \cite{sunder}
\begin{equation}\label{db}
\{f,g\}^*=\{f,g\}_{PB}-2i
\int\int dxdy\{f,C_{\scriptscriptstyle F}(x)\}_{PB}\delta(x-y)
\{C_{\scriptscriptstyle F}(y),g\}_{PB},
\end{equation}
where $f(z)$ and $g(z)$ are arbitrary phase--space functions.

Upon introducing the Dirac brackets one can eliminate 
$\psi^0$ (and $\bar\psi^0$ in the antiholomorphic sector) 
from the number of the physical variables of the model. 

Thus, within the course of the Hamiltonian reduction of the model we 
have encountered an unusual situation, namely, the appearance of the 
second--class constraint. Usually, when performing the Hamiltonian 
reduction of WZNW models, one restricts oneself to imposing first--class 
constraints \cite{bal,f,drs}. 

A peculiar feature of the case under consideration is that 
though superfield 
constraints \p{10} are a mixture of the first-- and second--class 
constraints, after the superconformal transformations \p{sc}
the first--class constraint component \p{cb} of \p{10} 
remains in the first class. 

The bosonic first--class constraint \p{1st} reflects the existence of 
the invariance of the model under the chiral transformations $g^+(Z)$ 
corresponding to the $E_+$ subalgebra of the $sl(2,{\bf R})$ algebra
\p{5}. We should stress that the parameter of these transformations is a 
chiral {\it superfield,} since the whole superfield constraint 
$J^-(Z)-1=0$ in \p{9} commutes (under \p{pbs}) with the superfield current 
$\Psi^-(Z)$ being the generator of $g^+(Z)$. A subtle point of the model 
at hand is that the presence of only one (bosonic) first--class 
constraint \p{cb} (while its superpartner \p{cf} is of the second class)
indicates that only the bosonic $g^+(Z)|_{\th=0}$--transformations 
form the real gauge symmetry of the model which can be subject to 
further gauge fixing. The fermionic $Dg^+(Z)|_{\th=0}$--transformations
(as the supersymmetry transformations 
\p{24.}--\p{+0}) are not fully fledged 
local transformations and cannot be used to reduce the number of 
physical degrees of freedom of the model. Note also that $\psi^+(z)$ 
transforms as a Goldstone field under $Dg^+(Z)|_{\th=0}$:
\begin{equation}\label{gold}
\d\psi^+(z)=-iDg^+(Z)|_{\th=0}.
\end{equation}

To analyse the superconformal structure of the physical sector of the 
model we impose an additional condition which fixes the 
$g^+(Z)|_{\th=0}$ gauge 
transformations and converts \p{cb} into a second--class constraint. 
Namely, as in the case of bosonic WZNW models \cite{bal} we impose the 
Drinfeld--Sokolov gauge \cite{ds}:
\begin{equation}\label{jo}
J^0|_{\th=0}=j^0=0.
\end{equation}

The constraint \p{jo} is invariant under superconformal transformations 
of the super current $J^0$ \p{6} provided the superconformal 
transformations are accompanied by a $g^+(Z)$--transformation whose 
leading component depends on superconformal parameters $\e(z)$, $\l(z)$ 
and supercurrent components. The explicit expression for 
$g^+(Z)|_{\th=0}$ is derived from the $J^0(Z)$ transformation law, which 
is obtained from \p{H}--\p{21} and \p{9}:
\begin{equation}\label{+}
{\hat J}^0=e^{-2\L}(J^0-iD\L\Psi^0)+e^{-\L-{1\over 2}D\L\Psi^-}\partial 
e^{-\L}-e^{-2\L}\partial e^{-{1\over 2}D\L\Psi^-} + 
e^{-2\L-D\L\Psi^-}g^+(Z).
\end{equation}
Having in mind that ${\hat J}^0|_{\th=0}=J^0|_{\th=0}=0$ we get for the 
infinitesimal transformations 
\begin{equation}\label{g+}
g^+(Z)|_{\th=0}=\partial^2\l(z)+i\e(\psi^+
+{1\over 2}\partial^2{\psi^-}+2i\psi^+\psi^-\partial{\psi^-}
-2\psi^-j^+)-{i\over 2}\partial\e\partial{\psi^-}.
\end{equation}
To treat the conditions \p{cb}, \p{jo} in the strong sense we should 
use them to construct new Dirac brackets with respect to which Eqs. 
\p{cb}, \p{jo} commute. For this to be achieved in a simplest way one 
should find a pair of second--class constraints, being a combination of  
\p{cb} and \p{jo}, which commute in a canonical way with respect to 
the Dirac brackets \p{db}. Their form turns out to be as follows:
$$
C^1_{\scriptscriptstyle B}=(j^--1)(1-{i\over 2}\psi^-\partial{\psi^-})=0,
$$
$$
C^2_{\scriptscriptstyle B}=j^0+{1\over 4}\partial C^1_{\scriptscriptstyle B}
(1+i\psi^-\partial{\psi^-})=0.
$$
\begin{equation}\label{cr}
\{C^1_{\scriptscriptstyle B}(z_1),C^2_{\scriptscriptstyle B}(z_2)\}^*=
\d(z_1-z_2), \qquad
\{C^1_{\scriptscriptstyle B}(z_1),C^1_{\scriptscriptstyle B}(z_2)\}^*=
\{C^2_{\scriptscriptstyle B}(z_1),C^2_{\scriptscriptstyle B}(z_2)\}^*=0.
\end{equation}
Using \p{cr} we construct new Dirac brackets
\begin{eqnarray}\label{ndb}
\{f,g\}^{**}& = &\{f,g\}^*-
\int\int dxdy\{f,C^1_{\scriptscriptstyle B}(x)\}^*\delta(x-y)
\{C^2_{\scriptscriptstyle B}(y),g\}^*\nn
&& +\int\int dxdy\{f,C^2_{\scriptscriptstyle B}(x)\}^*\delta(x-y)
\{C^1_{\scriptscriptstyle B}(y),g\}^*,
\end{eqnarray}
which allow one to treat the constraints \p{cr} in the strong sense.
Under the Dirac brackets \p{ndb} the fields $\psi^+(z)$, $\psi^{-}(z)$
and $j^+(z)$ 
have the following commutation properties:
\begin{equation}\label{psi}
\{\psi^+(z_1),\psi^+(z_2)\}^{**}=-{{ik}\over 2}\partial^2\d(z_1-z_2), 
\end{equation}
$$
\{\psi^+(z_1),\psi^-(z_2)\}^{**}=-ik\d(z_1-z_2),
$$
$$
\{\psi^-(z_1),\psi^-(z_2)\}^{**}=
\{j^+(z_1),\psi^+(z_2)\}^{**}=\{j^+(z_1),\psi^-(z_2)\}^{**}=0,
$$
\begin{equation}\label{j+}
\{j^+(z_1),j^+(z_2)\}^{**}={k\over 2} \partial^3\d(z_1-z_2)
+\d(z_1-z_2)\partial{j^+}(z_2)
-2\partial\d(z_1-z_2)j^+(z_2).
\end{equation}
In Eqs. \p{psi}, \p{j+} and below we restored the explicit dependence of 
quantities on the level $k$ of the WZNW model.
The commutation relations \p{psi} can be simplified even further if 
instead of $\psi^+$ one considers 
\begin{equation}\label{b}
b(z)=\psi^+(z)-{1\over 4}\partial^2{\psi^-}(z), 
\qquad c(z)={1\over k} \psi^-(z) 
\end{equation}
as the independent fermionic fields. Then $b(z)$ and
$c(z)$ are canonical conjugate free fields with spin 
${3\over 2}$ and $-{1\over 2}$, respectively:
\begin{equation}\label{bc}
\{b(z_1),c(z_2)\}^{**}=-i\d(z_1-z_2), \qquad
\{b(z_1),b(z_2)\}^{**}=\{c(z_1),c(z_2)\}^{**}=0.
\end{equation}
Eq. \p{j+} reads that $j^+(z)$ has the properties of the Virasoro 
stress tensor for the single bosonic (Liouville) mode \p{l}  
of our model and is indeed the same as in the purely bosonic case 
\cite{bal} (this can be checked by use of Eqs. \p{9}, \p{11} and taking 
into account the constraints):
\begin{equation}\label{tl}
T_{m}\equiv j^+= k[(\partial\phi)^2-\partial^2\phi].
\end{equation}
Thus in the classical case under consideration 
the full Virasoro stress tensor $T(z)$ and the superconformal 
current $G(z)$ constructed of $T_{m}(z)$, 
$b(z)$ and $c(z)$ have the 
following form:
$$
T(z)=T_{m}
-{3i\over {2}}b\partial c -{i\over {2}}\partial bc, 
$$
\begin{equation}\label{tg}
G(z)=ib+{i}cT_{m}-bc\partial c
-{k\over {4}}c\partial c\partial^2 c-ik\partial^2 c.
\end{equation}
The $n=1$ superconformal algebra of \p{tg} realized on the Dirac 
brackets \p{ndb} is
\begin{eqnarray}\label{sca}
\{T(z_1),T(z_2)\}^{**}&=&{k\over 2}\partial^3\d(z_1-z_2)
+\d(z_1-z_2)\partial{T}(z_2)
-2\partial\d(z_1-z_2)T(z_2), \nn
\{T(z_1),G(z_2)\}^{**}&=&\d(z_1-z_2)\partial T(z_2)
-{3\over 2}\partial\d(z_1-z_2)T(z_2) \nn
\{G(z_1),G(z_2)\}^{**}&=&{-2ki}\partial^2\d(z_1-z_2) +2\d(z_1-z_2)T(z_2).
\end{eqnarray} 
The central charge of the classical algebra is $c={6k}$.

The $n=1$ superconformal system described by \p{tg}, \p{sca} is a 
classical counterpart of the $b-c$ 
matter structure used in the framework of the universal string theory  
\cite{ber}. Hence, we can assert that in the case under 
consideration we deal with a particular example of an $n=(1,1)$, 
$D=3$ fermionic string in a physical gauge where two longitudinal bosonic 
modes are gauge fixed and the single (transversal) bosonic 
degree of freedom of the string is described by the Liouville mode
(compare with the case of the $N=2$, $D=3$ GS superstring of Section 3).
This gauge is alternative to the light--cone gauge of string theory, 
where also only the physical string modes remain.

We have thus established, at the classical level, links of the reduced 
supersymmetric $SL(2,{\bf R})$ WZNW model 
with the $N=2$, $D=3$ Green--Schwarz 
superstring on one hand and the $n=(1,1)$, $D=3$ fermionic string, 
carrying spin (${3\over 2},-{1\over 2}$) matter on the worldsheet, on the 
other hand. As to quantum relationship of the model 
to strings, the consideration 
of these points touches deep problems of comparing the results of 
quantization of constrained systems carried out before or after solving 
classical constraints and is beyond the scope of the present work.

So let us now turn to the quantization of the $n=1$ superconformal 
algebra \p{tg}, \p{sca} realized on the fields $\phi$, $b$ and $c$
without addressing to string connection.

To quantize this algebra we should take into account operator ordering
in \p{tg} and pass from the Dirac brackets to the quantum 
(anti)commutation relations, or to consider the quantum operator product 
expansion of the fields. We choose the second formalism, and as a result 
we get a realization of the quantum $n=1$ superconformal algebra which 
arose in \cite{bersh,ber}.

In the OPE language the super-Virasoro algebra we reproduce
is:
\begin{eqnarray}
\underline{T(z)T(0)} &=& {c/2\over z^4} + {2T\over z^2} + {T'\over z},
\nonumber\\
\underline{T(z)G(0)} &=& {(3/2) G\over z^2} + {G'\over z},\nonumber\\
\underline{G(z)G(0)} &=& {2c/3 \over z^3} + {2T\over z}.
\label{svira}
\end{eqnarray}
Here and in the following formulas the fields on the r.h.s. are
computed at $w=0$. The prime denotes for simplicity the
${\partial_z}$ derivative.\par
The OPE for the $b(z),c(z)$ field can be normalized as
\begin{eqnarray}
\underline{c(z)b(0)}&=& {1\over z},\nonumber
\end{eqnarray}
and  $j^+\equiv T_{m}(z)$  
satisfies the OPE for the bosonic matter stress tensor
with the quantum Liouville central charge $c_{m}= 6k+{{k+4}\over{k-2}}$
(see, for example, \cite{ishi}) 
and has regular OPE's with $b(z)$
and $c(z)$ (compare with Eqs. \p{j+}). 

The realization of the $n=1$ super--Virasoro
algebra in terms of the fields $T_{m}(z)$, 
$b(z)$ and $c(z)$ was constructed 
in \cite{ber}:
\begin{eqnarray}\label{superrepr}
T&=& T_{m} 
-{3\over 2} :bc': - {1\over 2} :b'c:  +{1\over 2}\partial^2(:cc':),
\nonumber\\
G &=& b+ :c T_{m}: +:bcc': -{{c_m-26}\over 24}:cc'c'': 
+ {{c_m-11}\over 6} c''
\end{eqnarray}
where dots denote the normal ordering,
and the quantum central charge is  $c= c_m-11$.
Notice the appearance of a new (the last) term in $T(z)$ and changing 
of coefficients of terms in $G(z)$ \p{superrepr} in 
comparison with the classical case \p{tg} due to the operator ordering.

To the $n=1$ superconformal algebra \p{svira} one can add the spin 
$(-{1\over 2})$  field $c(z)$ as the generator of the fermionic 
transformations \p{gold}. Then  we get  a nonlinear extension of the 
super--Virasoro algebra considered in \cite{sorin}.
\par
For completeness let us present an alternative realization
of the
super-Virasoro algebra which makes use of a field 
$V(z)$ (conventionally represented as $V(z)\equiv 
\partial\Phi(z)$) satisfying the free-field OPE
\begin{eqnarray}
\underline{V(z)V(0)}&=& -{1\over z^2}
\end{eqnarray}
together with the free fields $b(z)$ and $c(z)$.
It is given by
\begin{eqnarray}
T &=& -{1\over 2} :V^2: + q V' -{3\over 2} :bc':- {1\over 2} :b'c:,
\nonumber\\
G &=& {1\over 2} b - : V^2 c : - {2\over q} : V c': -{(1-2q^2)\over q} 
:V'c:,
\nonumber\\
&& +2 : cc'b:  + ({13\over 3} - {1\over q^2} - 4 q^2) :cc'c'':
-{2\over 3} (5 -6 q^2) c'',
\end{eqnarray}
where  $q=({{c_m-1}\over 12})^{1\over 2}$, and 
the central charge is the same as above. \par
Notice that in this case the stress-energy tensor $T(z)$ is the same as 
in the classical case \p{tg}, i.e. the standard
one given by the sum of two stress-energy tensors for the bosonic and 
fermionic free--field components, respectively.

\section{Hamiltonian reduction in the case of a general superaffine 
Lie algebra}

In this Section we shall discuss how to generalize the previous 
construction valid for the $sl(2)$ algebra to any given bosonic
algebra (or even superalgebra). A complete
analysis of this case
is postponed to future work, here we restrict ourselves
to outline the main features and basic ingredients. \par
To simplify discussion it is convenient to use the language
of the so-called
soldering procedure introduced by Polyakov in \cite{polya}, which
allows reproducing the results of the (more complete) Dirac analysis
performed above.\par
Let us start recalling that the $n=1$ chiral fermionic supercurrents
$\Psi^\pm (Z)$, $\Psi^0(Z)$ of the superaffine $sl(2)$
transform  under the infinitesimal
gauge transformations \\
parametrized by bosonic chiral superfields
$\epsilon^\pm(Z)$, $\epsilon^0(Z)$ as follows:
\begin{eqnarray}\label{P}
\delta\Psi^+ &=& 2(\epsilon^0 \Psi^+ -\epsilon^+ \Psi^0) + 
{1\over i}D\epsilon^+,
\nonumber\\
\delta\Psi^0 &=& \epsilon^+ \Psi^- -\epsilon^- \Psi^+ +
{1\over i}D\epsilon^0,\nonumber\\
\delta\Psi^- &=& 2(\epsilon^- \Psi^0 -\epsilon^0 \Psi^-) +
{1\over i}D\epsilon^-.
\end{eqnarray}
Using the Maurer Cartan equation \p{8}
we constructed from  $\Psi^\pm (Z)$, $\Psi^0(Z)$ the bosonic 
superfields $J^\pm$, $J^0$ \p{9}
\begin{eqnarray}
J^+ &=& D\Psi^+ -2i\Psi^0\Psi^+, \nonumber\\
J^0 &=& D\Psi^0 +i\Psi^-\Psi^+,\nonumber\\
J^- &=& D\Psi^- +2i\Psi^0\Psi^-.
\label{reconstruction}
\end{eqnarray}
By construction they obey the following $sl(2)$ transformation properties
\begin{eqnarray}\label{J}
\delta J^+ &=& 2(\epsilon^0 J^+ -\epsilon^+ J^0) +
\partial \epsilon^+,\nonumber\\
\delta J^0 &=& \epsilon^+ J^- -\epsilon^- J^+ +
\partial \epsilon^0,\nonumber\\
\delta J^- &=& 2(\epsilon^- J^0 -\epsilon^0
J^-) +\partial \epsilon^-.
\end{eqnarray}
They correspond to a trivial supersymmetrization of the $sl(2)$ affine 
algebra obtained by replacing the ordinary currents and parameters
with the chiral superfields $J^i$ and $\epsilon^i$.
Since the ring structure properties of the superfields are the same as
for the ordinary fields it is clear that any bosonic theory can be
trivially supersymmetrized that way. In general it leads to 
uninteresting models. What makes things
different here is the presence
of the set of equations \p{reconstruction}, 
which allow expressing the superfields $J^i$ in 
terms
of the basic superfields $\Psi^i$.

The constraint $J^- -1 = 0$ and the gauge fixing condition 
${J^0}_{|\theta = 0} =0$ can be consistently
imposed as discussed in Section 2, 4 and 5. Then the only independent 
transformations which remain in \p{P} and \p{J} are parametrized by the 
bosonic component of $\epsilon^-(Z)$ which coincides with the conformal 
transformation parameter $\l(z)$ in \p{conb}--\p{confor}, the fermionic 
component of $\e^-(Z)$ which is connected with supersymmetry parameter 
$\e(z)$ in \p{24.}--\p{+0} through the relation 
${1\over i}D\e^-|_{\th=0}=\e(1+i\psi^-\partial\psi^-)-{1\over 2}\partial 
\l\psi^-$,
 and an additional fermionic transformation parameter 
$D\e^+|_{\th=0}$ (compare with \p{gold}). 

This reasoning can be applied to get information about
the reduction of more general $n=1$ superaffine algebras.

We wish to stress that the bosonic superfields $J^{\pm}$ associated
to the roots of the $sl(2)$ algebra are obtained, as a consequence
of the Maurer-Cartan equations \p{8}, by applying to $\Psi^{\pm}$
a fermionic derivative
covariant with respect to the superaffine $U(1)$ subalgebra of 
$sl(2)$. 
 \footnote{Covariant derivatives to analyze
algebra structures have been introduced in 
\cite{top0} and further employed in 
\cite{top2} in connection with integrable hierarchies; fermionic 
covariant derivatives have been introduced in \cite{top3}.}. 
This property obviously
takes place in the case of a generic bosonic algebra. Let $\Psi^{\pm j}(Z)$ 
denote generic superfield currents associated to a root system of 
an $n=1$ superaffinization
of a given bosonic simple Lie algebra ${\cal G}$, 
and $\Psi^{0,\a}(Z)$ denote the supercurrents associated to the Cartan 
subalgebra (here $\a=1,...,r$, and $r$ is the rank of ${\cal G}$). 
The supercurrents
$\Psi^{\pm j}(Z)$ associated to the simple and nonsimple roots
are covariant w.r.t. the $n=1$ superaffine subalgebra ${\hat 
{U}(1)}^r$ of ${\cal G}$, i.e. they satisfy the following Poisson brackets:
\begin{eqnarray}
\{\Psi^{0,\a}(Z), \Psi^{\pm j } (W)\}&=& \pm q^{\a,j}\Psi^{\pm j}(W)\delta 
(Z-W).\nonumber
\end{eqnarray}
where $q^{\a,\pm j}= \pm q^{\a,j}$ is a set of $U(1)$--charges. When 
$+j=\b$ 
labels a simple root,  $q^{\a,\b}$ coincides with the Cartan matrix
$K_{\a\b}$.

The bosonic superfields $J^{\pm j}$ associated to the roots
are constructed from $\Psi$
as 
\begin{equation}\label{tok}
J^{\pm j} = {\cal D} \Psi^{\pm j} + i\sum_{ k, l} 
{c^{\pm j}}_{k,l} \Psi^{k}\Psi^{l},
\end{equation} 
where the covariant derivative is 
uniquely determined  as ${\cal D} \equiv D 
+\sum_{\a=1,...,r} q^{\a,j} \Psi^{0,\a}$  by the requirement that
$J^{\pm j}$ satisfy the Maurer--Cartan equations and 
are covariant with the charge $q^{\a,j}$:
\begin{eqnarray}
\{\Psi^{0,\a}(Z),J^{\pm j}(W)\} &=& \pm q^{\a,j} J^{\pm j} (W)\delta(Z-W)
\nonumber
\end{eqnarray}
The constants ${c^{\pm j}}_{k,l}$ in \p{tok}
are the Lie algebra structure constants; they are  
non-vanishing only
for those integral values of $k,l$ for which
$q^{\a, k} + q^{\a,l} = \pm q^{\a, j}$ for any $\a$.

Similarly the bosonic supercurrents 
$J^{0,\a}$ associated to the Cartan subalgebra
are given by
\begin{eqnarray}
{J^{0,\a}} &=& D\Psi^{0,\a} + i\sum_j {c^\a}_{+j,-j} \Psi^{+j}\Psi^{-j}
\nonumber
\end{eqnarray}
(here again the coefficients ${c^\a}_{+j,-j}$ 
are Lie algebra structure constants).

Now the set of constraints for the so-called abelian reduction can be 
consistently imposed as in the supersymmetric $sl(2)$ and 
purely bosonic case: the set of
superfields $J^{-1, i}$ \p{tok} 
associated to the simple (negative) roots are put
equal to one: $J^{-\a} = 1$, while the remaining generators associated
to lower (negative) roots are set equal to $0$, as well as the
bosonic components of the Cartan supercurrents $J^{0,\a}$.        
The soldering procedure and the Dirac analysis for these constraints 
can be made as in the $sl(2,{\bf R})$ case. As a result
superaffine transformations are converted into extended super--Virasoro 
transformations and one may
get an $n=1$ supersymmetric extension of the bosonic ${\cal W}$
algebra realized on the bosonic ${\cal W}$ algebra 
generators and $r$ pairs of free fermions.
 
It deserves mentioning that the standard supersymmetrization of a 
bosonic ${\cal W}$ algebra is carried 
out in two steps (see \cite{sorba2}),
namely, 
by identifying a suitable superalgebra and an embedding 
of $osp(1|2)$ or $sl(1|2)$ into it. 
In our approach we perform the supersymmetrization of 
the ${\cal W}$ algebra by direct use of the same ingredients as in the 
bosonic case: the classical Lie algebra as a base and   
a given embedding of $sl(2)$ into the former (the principal
embedding for abelian theories).
Because of the presence of 
spin $({3\over 2},-{1\over 2})$ $b-c$ 
systems
such a realization of a super--${\cal W}$ 
algebra differs in field contents
from the standard ones based on superalgebras described by the simple 
fermionic root system \cite{sorba2}.
A reduction procedure alternative to ours was used in \cite{bersh} 
to get $n=2$ superconformal 
$b-c$ structures and realizations of super-$W_n$ algebras from the 
$Sl(n|n-1)$ superalgebras in a Gauss decomposition containing simple 
bosonic roots.
A class of supersymmetric ${\cal W}$ algebras realized with the use of
fermionic $b-c$ systems has also been considered in \cite{sorin} 
and references therein.

Super--Toda equations obtained with the procedure under consideration
are a generalization of Eqs. \p{12}, \p{13}:
\begin{equation}\label{12g}
\bar DD\Phi_\a=e^{\sum_\b K_{\a\b}\Phi_\b}\bar\Psi^{+\a}\Psi^{-\a}, 
\qquad \bar 
D\Psi^{-\a}=0=D\bar\Psi^{+\a};
\end{equation}
\begin{equation}\label{13g}
D\Psi^{-\a}+\sum_{\b}K_{\a\b}\Psi^{0,\b}\Psi^{-\a}=1, \qquad
\bar D\bar\Psi^{+\a}+\sum_{\b}K_{\a\b}\bar\Psi^{0,\b}\bar\Psi^{+\a}=1.
\end{equation}
In Eqs. \p{12g}, \p{13g} there is no summation over $\a$; 
$\Phi_\a(Z,\bar Z)$ are the Cartan subalgebra superfields of the Gauss 
decomposition of the group element
$$
G(Z,\bar Z)=e^{\sum\b^\a E_{+\a}}e^{\sum\Phi_{\a}H_\a}e^{\sum\g^\a 
E_{-\a}}
$$
and (with taking into account the constraints imposed)
$$
\Psi^{0,\a}={1\over i}D\Phi_{\a}+\b^\a\Psi^{-\a}, \qquad
\bar\Psi^{0,\a}=i\bar D\Phi_{\a}+\g^\a\bar\Psi^{-\a}~~~({\rm no~summation 
~over~\a}).
$$
Notice that because of the form of $\Psi^{0,\a}, ~\bar\Psi^{0,\a}$ and 
the summation over $\b$ in \p{13g} the constraints \p{13g}
do not reduce to relations which contain only $D\Phi_{\a}$ (in contrast 
to Eqs. \p{10}, \p{13}). A consequence of this is that, unlike in the 
simplest $Sl(2,{\bf R})$ case, fermions do not decouple from bosons in 
the component form of Eqs. \p{12g}. (At least we have not managed to 
find a local field redefinition which would allow one to completely 
decouple fermions).

The detailed studying of super--Toda models of this kind may be of 
interest because of unusual supersymmetric properties, their connection 
with standard super--Toda models, and of possible 
physical applications.
 
The Hamiltonian reduction procedure outlined herein can also be applied 
to those superalgebras which always contain in their root 
decomposition bosonic simple roots. In this case one should impose 
a ``mixed" type of constraints, namely, 
standard ones associated to fermionic simple roots and nonlinear ones
for the bosonic roots as discussed in this paper. 
It is very likely that the models derived this way 
have a nontrivial
interaction between the bosonic and fermionic sector. 
As far as we know, without explicit breaking supersymmetry 
\cite{hollo}, superalgebras of this kind have not been 
involved yet into
the production of super--Toda models.
The complete classification 
of superalgebras and their Dynkin diagrams
can be found in \cite{sorba}. From this paper we learn that, for instance,
up to rank $3$, there 
exist $6$ superalgebras which contain bosonic simple roots in any root 
decomposition.
The rank $2$ superalgebra of this kind is $Osp(1|4)$, and the rank $3$ 
superalgebras are $Osp(5|2)$, $Osp(2|4)$, $Osp(1|6)$,
$ Sl(1|3)$ and $G(3)$. \par
Apart from these systems which deserve to be analyzed
in detail,
there is another problem which is worth studying. Many 
superalgebras admit different variants of root decomposition. 
Even superalgebras
which are expressible in terms of only fermionic simple roots  can be
expressed, in another basis, in terms of a Dynkin diagram
involving bosonic simple roots (the simplest superalgebra having this
property is $Sl(1|2)$). It is interesting to understand
possible relations, if any, between super--Toda models derived
from different
presentations of the same superalgebra. 

\section{Conclusion and discussion}
We have studied the Hamiltonian 
reduction of the $n=(1,1)$ supersymmetric WZNW models having a classical 
(bosonic) group as target space.

Since the simple roots of the corresponding Lie algebra are bosonic 
the constraints are to be imposed on associated bosonic supercurrents 
constructed out of the basic fermionic supercurrents in a nonlinear way 
prescribed by the Maurer--Cartan equation. The constraints thus obtained 
are a mixture of bosonic first--class and fermionic second--class 
constraints. This makes difference between the Hamiltonian reduction of the 
models considered above and the conventional Hamiltonian reduction of 
the bosonic WZNW models \cite{bal,f} and the supersymmetric WZNW models 
based on superalgebras admitting realization in terms of fermionic 
simple roots only \cite{drs}. In the latter case all the constraints 
imposed are of the first class.

Hamiltonian reduction results in shifting the conformal spins of  
bosonic as well as of fermionic current fields. As a result we obtained 
the fermionic $b-c$ system with spin $3\over 2$ and $-{1\over 2}$. 
Supersymmetry
transforming these fields is realized nonlinearly and is spontaneously 
broken.

The supersymmetric Toda systems obtained by reducing the WZNW models 
with bosonic groups contain a 
free left-- right--chiral fermion sector. 
In the simplest case of the $Sl(2,{\bf R})$ group the fermionic sector 
completely decouples from the bosonic Liouville equation. While in the 
case of super--Toda models based on higher rank bosonic groups fermions 
contribute to the r.h.s. of the bosonic equations.
Thus the supersymmetric 
generalization of the bosonic Toda--models obtained this way is 
alternative to the conventional one (see e.g.\cite{top}) 
which involves nontrivial 
interactions in both the bosonic and fermionic sector and where all fermionic
fields have positive conformal spin. It seems of interest to analyse in 
detail the relationship between these two supersymmetric versions.

An important problem which has remained unsolved yet is the construction 
of a gauged WZNW action from which one can directly get the constraints 
as equations of motion of auxiliary gauge fields. As we have already 
mentioned in the main text, the problem is caused by the nonlinear 
form of the constraints which, in addition, contain bosonic coordinate 
derivatives of group--valued superfields $G(Z)$, 
while the WZNW action \p{1} is constructed 
with the use of only Grassmann supercovariant derivatives. The cause
of the problem seems akin to the problem of constructing superfield 
actions for $n$--extended WZNW models. 

Possible directions of generalizing results obtained are 
following.

The detailed analysis of the Hamiltonian reduction of the $Sl(2,{\bf 
R})$ WZNW model at the classical level 
revealed its connection with $N=2$ Green--Schwarz 
superstrings and $n=(1,1)$ fermionic strings propagating in 
flat $D=3$ space--time. The extension to the quantum level and
the analysis of links of the present model
with the standard $n=(1,1)$, D=3 Neveu--Schwarz--Ramond string whose 
fermionic matter fields have conformal spin ${1\over 2}$ require
additional study since it involves nontrivial field redefinitions when 
passing from one superstring formulation to another (see, for example
\cite{vz,berk,berko}).

Studying higher-dimensional target spaces seems also of interest. 
As we know from the doubly supersymmetric 
approach (see \cite{bpstv} and references therein), in 
D=4 flat superspace--time, whose structure 
group is $Sl(2,{\bf C})$, Green--Schwarz superstrings are described by 
an $n=2$ worldsheet supersymmetry. Hence we can associate to these strings 
an $n=2$ supersymmetric $Sl(2,{\bf C})$ WZNW model \cite{h}
appropriately reduced the way considered above. For the  $N=2$, $D=4$ 
superstring this should result in a supersymmetric integrable system
\footnote{A relation of four--dimensional n=2,4 supersymmetric string 
backgrounds to integrable models has been already considered, for 
instance, in \cite{cl}} 
which describes the physical modes of the classical superstring and 
which consists of a bosonic Liouville system and two pairs 
of left-- right--moving free fermions. As in the D=3 case, one may also 
expect a connection with an $n=(2,2)$ superconformal system 
corresponding to an $n=(2,2)$ fermionic string \cite{ber}.

Next, more complicated, step is to perform the Hamiltonian reduction of 
supersymmetric WZNW models based on $SO(1,5)$ and $SO(1,9)$ group which 
correspond, respectively to $N=2$ Green--Schwarz superstrings in $D=6$ 
and $D=10$ space--time with $n=(4,4)$ and $n=(8,8)$ supersymmetries on 
the worldsheet, or at least with $n=(2,2)$ manifest worldsheet 
supersymmetries, and to establish a connection with results of Refs. 
\cite{ber,berk,berko} on Neveu--Schwarz--Ramond and Green--Schwarz 
superstrings. Another words one should study 
a generalization of the Hamiltonian reduction
procedure considered here to the case of WZNW 
models with extended supersymmetry. 

Though we have talked of $N=2$,  Green--Schwarz superstrings in D=3,4,6 
and 10 dimensions and corresponding $n=(D-2, D-2)$ superconformal models 
the discussion above is also valid for $N=1$ 
Green--Schwarz superstrings and chiral (heterotic) $n=(D-2,0)$ 
superconformal models.

Another direction of study, which may turn out the most interesting from 
a mathematical physics point of view, is to apply the Hamiltonian 
reduction procedure to constrain supersymmetric WZNW models based on 
superalgebras with fermionic and bosonic simple roots in any Dynkin 
diagram \cite{sorba}.
The simplest example is the $OSp(1|4)$ superalgebra whose root 
decomposition contains one bosonic and one fermionic simple root.
A potentially promising investigation concerns other
supergroups of this kind, the three--rank exceptional 
supergroup $G(3)$ 
and the four--rank exceptional supergroup $F(4)$. It is likely
\cite{ketov} that their reductions  
are related respectively to an $n=7$ and $n=8$ supersymmetric 
extension of the conformal algebra.

\bigskip
{\bf Acknowledgements.}\\
~\par
The authors are grateful to I. Bandos, F. Bastianelli, 
E. Ivanov, S. Ketov, S. Krivonos,
A. Pashnev, P. Pasti, A. Sorin, M. Tonin and D. Volkov 
for interest to this work and valuable discussion.

\end{document}